\documentstyle[12pt,openbib]{article}
\title{A Proposal for Solving the ``Problem of Time"
in Canonical Quantum Gravity}
\author{Robert M. Wald\\
         {\it University of Chicago}\\
         {\it Enrico Fermi Institute and Department of Physics}\\
         {\it 5640 S. Ellis Avenue}\\
         {\it Chicago, Illinois 60637-1433}}
\date{}

\begin{document}
\maketitle
\begin{abstract}
The ``problem of time" in canonical quantum gravity refers to
the difficulties involved in defining a Hilbert space structure
on states -- and local observables on this Hilbert space -- for
a theory in which the spacetime metric is treated as a quantum
field, so no classical metrical or causal structure is present
on spacetime. We describe an approach -- much in the spirit of
ideas proposed by Misner, Kuchar and others -- to defining
states and local observables in quantum gravity which exploits
the analogy between the Hamiltonian formulation of
general relativity and that of a relativistic particle. In the case of
 minisuperspace models, a concrete theory is obtained which
 appears to be mathematically and physically viable, although
 it contains some radical features with regard to the presence of
 an ``arrow of time". The viability of this approach in the case of
 infinitely many degrees of freedom rests on a number of fairly
 well defined issues, which, however, remain
unresolved. As a byproduct of our analysis, the theory of a
relativistic particle in curved spacetime is developed.

{\bf PACS \#: } 04.60.+n, 04.20.-q, 03.70.+k
\end{abstract}
\newpage

	A key issue which arises in any attempt to obtain a quantum
theory of general relativity -- or, more generally, any theory in which
the spacetime metric is treated as a quantum observable -- is how to
formulate a local field theory without the presence of any classical
metrical or causal structure on spacetime. This issue is faced most
directly in ``non-perturbative" approaches to quantization, such as
the canonical approach.

	As is well known, classical general relativity admits a
Hamiltonian formulation, but this formulation possesses constraints
which are closely analogous to those occurring in the theory of a
relativistic particle. The canonical approach to formulating a
quantum theory corresponding to general relativity begins with this
Hamiltonian formulation \cite{ft1}, in which the role of
configuration variable is played by a Riemannian metric, $h_{ab}$, on
a three-dimensional manifold, $\Sigma$, and the conjugate
momentum, $\pi^{ab}$, has the interpretation of being directly
related to the extrinsic curvature of $\Sigma$ in the classical
spacetime obtained by evolving this initial data. The states of the
quantum theory are then taken to be wavefunctionals,
$\Psi [h_{ab}]$, of the metric on $\Sigma$, and the constraints of the
classical theory are imposed as conditions on $\Psi$. The
``momentum constraints" imply that $\Psi$ is spatial diffeomorphism
invariant, (i.e., that it depends only upon the three-geometry), and
they can be ``solved" by taking $\Psi$ to be a function on
``superspace", the manifold of three-geometries.
The Hamiltonian constraints give rise to the Wheeler-DeWitt
equations on $\Psi$. Note that I use the plural here to stress that
there is an infinite family of Hamiltonian constraints -- one for each
choice of ``lapse function" -- and a correspondingly infinite family of
Wheeler-DeWitt equations.

The difficulties with the canonical
approach arise when one attempts to impose a Hilbert space
structure on the allowed state vectors, and when one attempts to
define operators on this Hilbert space for observables of interest
such as $h_{ab}$ and $\pi^{ab}$. In particular, it is far from clear
what in the theory should play the role of a ``Heraclitian time
variable" \cite{uw}, which ``sets the conditions" for determining
probabilities for the values of the dynamical variables. We refer the
reader to \cite{ku1} (see also section 1 of \cite{uw}) for
comprehensive review of the various approaches that have been
taken to this ``problem of time" and the serious difficulties which
these approaches have encountered.

	The purpose of this paper is to describe a proposal for defining
Hilbert space structure and observables in canonical quantum
gravity. The basic ideas involved in this proposal are not new; they
appear in the work of Misner \cite{mis}, Kuchar (see particularly
\cite{ku2}), and others. However, it now appears that it may be
possible to overcome two potentially serious obstacles to the
implementation of these ideas. We shall focus attention upon the case
of minisuperspace models, where regularization issues do not arise,
and the proposal can be given a concrete form. Issues related to the
generalization of this proposal to the case of infinitely many degrees
of freedom will be addressed at the end of this paper.

	For definiteness, we focus attention upon
``class A" Bianchi cosmological
models, whose only matter content is a homogeneous scalar field
$\phi$. (Class A Bianchi
Lie algebras are the three-dimensional ones in which
the structure tensor takes the form
${c^a}_{bc}$ = $M^{ad} \epsilon_{dbc}$ with $M^{ad}$ symmetric and
$\epsilon_{dbc}$ totally antisymmetric; restriction to class A models
is made in order to assure the existence of a Hamiltonian formulation
of the minisuperspace dynamics.) For such a model, minisuperspace
is a 4-dimensional manifold, with three parameters characterizing
the spatial geometry and one giving the value, $\phi$, of the
homogeneous scalar field. We choose the parametrization
$(\alpha, \beta_+ , \beta_-)$ of the spatial metric introduced by
Misner (see, e.g., box 30.1 of \cite{mtw}) where, in essence,
$\alpha$ measures the volume of the universe, and
$(\beta_+, \beta_-)$ measure the spatial anisotropy. With
appropriate re-scalings of variables, the superhamiltonian takes the
form,
\begin{equation}
H = - {p_{\alpha}}^2 + {p_{\beta_+}}^2 +
{p_{\beta_-}}^2 + {p_{\phi}}^2 +
\exp(4\alpha) V_{\beta}(\beta_+,\beta_-) +
\exp(6\alpha) V_{\phi} (\phi)
\label{sh}
\end{equation}
where $(p_{\alpha}, p_{\beta_+}, p_{\beta_-}, p_{\phi})$ are the
momenta canonically conjugate to $(\alpha, \beta_+, \beta_-, \phi)$
and the ``potential", $V_{\beta}$, depends upon the choice of Lie group.

	In the canonical approach to the quantization of this model, it is
generally agreed that in the ``metric representation", the Hamiltonian
constraint should be enforced by requiring the state vector,
$\Psi (\alpha, \beta_+, \beta_-, \phi)$ to satisfy the Wheeler-DeWitt
equation
\begin{equation}
\hat{H} \Psi = 0
\label{wdw}
\end{equation}
where $\hat{ H}$ is obtained from $H$ by replacing the classical
momenta with the corresponding differentiation operators. Thus,
more explicitly, $\Psi$ satisfies
\begin{equation}
[G^{AB} \nabla_A \nabla_B - \exp(4\alpha) V_{\beta} -
\exp(6\alpha) V_{\phi}] \Psi = 0
\label{wdw2}
\end{equation}
where the DeWitt supermetric, $G_{AB}$, is just a flat, Lorentz
signature metric in the ``global, inertial, superspace coordinates"
$(\alpha, \beta_+, \beta_-, \phi)$. However, serious difficulties occur
when one attempts to define a Hilbert space structure, ${\cal H}$, on
the states and to define self-adjoint operators on ${\cal H}$
representing observables of interest. Our aim, now, is to overcome
these difficulties.

	The superhamiltonian (\ref{sh}) has the same mathematical
form as that of a relativistic particle in a (time- and space-dependent)
potential, and, correspondingly, the Wheeler-DeWitt equation
(\ref{wdw2}) has the mathematical form of a Klein Gordon equation
in an external potential. This analogy is not superficial, nor is it
special to the particular class of models explicitly considered here:
The presence of a constraint in the theory of a
relativistic particle traces its origin to the treatment of the time
coordinate of the particle as a dynamical variable, and the origin of
the Hamiltonian constraint of general relativity can be given a
similar interpretation. Furthermore, the Lorentz signature of
$G_{AB}$ is not an artifact of our model; in any homogeneous model,
$G_{AB}$ will have Lorentz signature (with the motion in superspace
associated with the conformal scaling of the 3-metric
being ``timelike") provided only that the kinetic energy
terms of the matter fields enter with the usual sign. Thus, it seems
natural to seek guidance for the definition of ${\cal H}$ and
observables on ${\cal H}$ from the theory of a
relativistic particle in a
curved spacetime and/or external potential.

	However, the theory of a relativistic particle in a
non-stationary curved spacetime or external potential is plagued by
some well known difficulties. Historically, these difficulties were
cured by passing to a ``second quantized" theory, i.e., by changing the
nature of the theory to that of a quantum theory of a field (with the
value of the field at spacetime events playing the role of the primary
observables of the theory) rather than a theory of a particle (where
the primary observables are the spatial position and momentum of
the particle). This step is well justified physically: It appears that
nature truly is described at a fundamental level -- or, at least to the
 level we currently are able to probe -- by quantum field theory.
However, since canonical quantum gravity already is structured as a
field theory, an analogous step here would correspond to a ``third
quantized" theory, in which the primary
observables presumably would become the value of the Wheeler-DeWitt
wavefunction at points of superspace. It seems difficult to imagine
how such ``observables" might be measured, or how such a theory
could be interpreted so as to give predictions about what one
ostensibly is interested in trying to describe in quantum gravity,
namely local metrical structure as determined by observers in
spacetime (as opposed to, say, an S-matrix describing the scattering
of multiple universes). Thus, I shall not pursue this avenue of
approach here.

	As I shall now explain, it is possible to give a mathematically
consistent, interpretable theory of a relativistic particle in a
non-stationary spacetime or external potential. However, this theory
suffers from two serious defects. Remarkably, these defects do not
appear to be impediments to the viability of an analogous quantum
theory for our minisuperspace models. As discussed briefly at the end of
this paper, the issue of whether these ideas can be extended to
provide a mathematically consistent and physically interpretable
quantum theory of gravity in the case of infinitely many
degrees of freedom remains open.

	Let $(M, g_{ab})$ be an arbitrary globally hyperbolic
spacetime, on which there is prescribed a
(possibly time- and space-dependent)
``external potential" $V$, and consider a ``relativistic
particle" on this spacetime, with classical Hamiltonian, $h$, given by
\begin{equation}
h = g^{ab} p_a p_b +  V
\label{rp}
\end{equation}
and where $h$ is constrained to vanish. We wish to construct a
quantum theory in which states are represented by (complex)
wavefunctions, $\Phi$, on spacetime and where the constraint
$h = 0$ is imposed by requiring $\Phi$ to satisfy the Klein-Gordon
equation,
\begin{equation}
g^{ab} \nabla_a \nabla_b \Phi -  V \Phi = 0
\label{kge}
\end{equation}
Now, the vector space of complex solutions to (\ref{kge}) possesses
the natural conserved, non-positive-definite, inner product,
\begin{equation}
<\Phi_1, \Phi_2>_{KG} = - i \Omega (\bar{\Phi}_1, \Phi_2)
\label{kg}
\end{equation}
where the (real) symplectic product, $\Omega$, is given by
\begin{equation}
\Omega (\Phi_1, \Phi_2) = \int_{\Sigma} [\Phi_2 \nabla_a \Phi_1
- \Phi_1 \nabla_a \Phi_2] d \Sigma^a
\label{sf}
\end{equation}
and $\Sigma$ is any Cauchy surface. The aim is to define the Hilbert
space of states, ${\cal H}$, by, in effect, choosing a suitable subspace
of complex solutions on which $< , >_{KG}$ is positive definite. As
discussed in detail in \cite{wa1}, this can be done by specifying a
real inner product, $\mu$, on the vector space of real, smooth
solutions to (\ref{kge}) with initial data of compact support, such
that,
\begin{equation}
\mu(\Phi_1, \Phi_1) = {}^{l.u.b.}_{\Phi_2 \neq 0}
\frac{|\Omega (\Phi_1, \Phi_2)|^2}{4 \mu(\Phi_2, \Phi_2)}
\label{mu}
\end{equation}
In the case of a stationary spacetime with a time-independent
potential $V > 0$, a natural choice of $\mu$ exists \cite{ak}, which
corresponds to taking ${\cal H}$ to be the subspace of positive
frequency solutions. This construction also can be used to define a
natural choice of $\mu$ if the spacetime and potential are merely
asymptotically stationary in the past or future. However, in the
absence of time translation symmetry, there does not appear to be
any natural choice of $\mu$. The difficulty is {\it not} that $\mu$'s
satisfying eq.(\ref{mu}) fail to exist -- there always exist a wide class
of inner products satisfying eq.(\ref{mu}) -- but rather that no
solution seems in any way uniquely ``distinguished". Indeed, note
that in the case where the spacetime and potential are
asymptotically stationary in both the past and the future, there will
be two ``distinguished" choices of $\mu$, which, in general, will
differ. The lack of a natural choice of $\mu$ -- and, thereby, of
${\cal H}$ -- is the first serious deficiency of the theory of a
relativistic particle in curved spacetime \cite{ft2}.

	Nevertheless, suppose an inner product, $\mu$, satisfying
eq.(\ref{mu}) has been chosen. I shall assume, in addition, that the
Hilbert space of solutions, ${\cal H}$, determined by this $\mu$
satisfies the following further properties: Given any Cauchy surface,
$\Sigma$, let ${\cal D}$ denote the subspace of ${\cal H}$ comprised
of $C^1$ solutions whose restriction to $\Sigma$ lies in
$L^2 (\Sigma)$. Now, any $\Phi \in {\cal D}$ is uniquely
characterized by this restriction, since if $\Phi$ and $\Phi '$ had the
same restriction to $\Sigma$, then the norm of $\Phi - \Phi '$ would
vanish by eqs.(\ref{kg}) and (\ref{sf}).
Consequently, ${\cal D}$ also may be viewed
as a subspace of the Hilbert space $L^2 (\Sigma)$. I shall assume that
${\cal D}$ is dense both as a subspace of ${\cal H}$ and as a subspace
of $L^2 (\Sigma)$. In addition, I shall assume that if $\{\Phi_n\}$ is
any sequence in ${\cal D}$ which converges in both ${\cal H}$ and
$L^2 (\Sigma)$, then the limit in ${\cal H}$ is non-zero if and only if
the limit in $L^2 (\Sigma)$ is non-zero. I believe that it is likely that
these assumptions could be proven to hold in the case (relevant for
our considerations below) of a spacetime which is asymptotically
stationary in the past, with $\mu$ chosen as described above, but I
have not attempted to investigate this issue carefully.

	Our aim, now, is to define operators on ${\cal H}$
corresponding to the position and momentum of the particle at an
arbitrary time. More precisely, given any Cauchy surface (i.e.,
``time"), $\Sigma$, for each function $f : \Sigma \rightarrow I\!\! R$
we wish to obtain a self-adjoint operator
$\hat{f} : {\cal H} \rightarrow {\cal H}$ whose spectral resolution
yields the probability distribution for the value of $f$ at the position
of the particle on $\Sigma$. Similarly, given any vector field, $v^a$,
on $\Sigma$ with complete orbits, we wish to obtain a self-adjoint
operator, $\hat{v}$, which can be interpreted as the infinitesimal
generator of translations of states by the diffeomorphisms on
$\Sigma$ generated by $v^a$. For flat spacetime with no potential
and for $\Sigma$ chosen to be a flat hyperplane, such operators were
defined by Newton and Wigner \cite{nw} for the case where $f$ is a
Cartesian coordinate on $\Sigma$ and $v^a$ is a Euclidean
translation. Thus, we wish to generalize this construction to curved
spacetimes, to arbitrary choices of $\Sigma$, and to general choices
of $f$ and $v^a$.

	On the Hilbert space $L^2 (\Sigma)$, there is a standard
prescription (see, e.g., appendix C of \cite{ash}) for defining position
and momentum operators of the sort described above. However, the
relevant Hilbert space here is ${\cal H}$, and direct application of the
 $L^2$-operators to the initial data associated with solutions in
${\cal H}$ will not, in general, even yield maps of ${\cal H}$ into
itself, no less yield self-adjoint operators on ${\cal H}$. Nevertheless,
we can proceed in the following manner, which generalizes a
construction of Kuchar (unpublished) for static hypersurfaces in
static spacetimes.

	Let $\Sigma$ be a Cauchy surface, let ${\cal D}$ be defined as
above, and let $\bar{{\cal D}}$ denote the Cauchy completion of
${\cal D}$ in the norm $||$ $||_{{\cal H}} + ||$ $||_{L^2 (\Sigma)}$. By our
assumptions above, we may view $\bar{{\cal D}}$ as a (dense)
subspace of ${\cal H}$, on which the $L^2$ inner product is well
defined and positive definite. We view the $L^2$ inner product as a
quadratic form $q$ on ${\cal H}$ with form domain $\bar{{\cal D}}$.
Since $q$ is closed and positive definite, it follows from general
results on quadratic forms (see, e.g., section 8.6 of \cite{rs})
together with the square root lemma that there exists a unique,
positive self-adjoint operator $A : {\cal H} \rightarrow {\cal H}$ with
domain $\bar{{\cal D}}$ such that for all
$\Phi_1, \Phi_2 \in \bar{{\cal D}}$, we have,
\begin{equation}
<\Phi_1, \Phi_2 >_{L^2}  =  <A \Phi_1, A \Phi_2 >_{{\cal H}}
\label{l2}
\end{equation}
Since $ker (A) = 0$, it follows that $A$ has dense range in ${\cal H}$.
Now view $A$ as a linear map from
$\bar{{\cal D}} \subset L^2 (\Sigma)$ into ${\cal H}$. By eq.
(\ref{l2}), this map preserves inner products. Since both the range
and domain of $A$ are dense, it follows that $A$ uniquely extends to
a unitary map $U : L^2 (\Sigma) \rightarrow {\cal H}$. We now
simply use this unitary correspondence to ``transport" to ${\cal H}$
the position
and momentum operators defined on $L^2 (\Sigma)$.
This prescription
reproduces the Newton-Wigner operators in the case considered
by them \cite{nw}.

	Thus, having chosen a $\mu$ which satisfies eq.(\ref{mu}) and
our additional assumptions, we obtain a theory in which for any
choice of ``time" (i.e., Cauchy surface), we have well defined
operators describing the position and momentum of the particle at
that time. These operators satisfy the usual commutation relations,
and I would not anticipate any difficulties with the existence of a
semiclassical limit of the theory in which the dynamics agrees closely
with that of a classical relativistic particle. However, there is a
serious deficiency of the theory: The exact quantum dynamics will
not respect the causal structure of the underlying spacetime, i.e., a
particle which, with unit probability, lies within a region $R$ on the
Cauchy surface $\Sigma$ will not, in general, be localized to within
the causal future of $R$ on a later Cauchy surface $\Sigma'$. (Indeed,
this phenomenon is well known to occur even for the theory
obtained with the Newton-Wigner observables in flat spacetime.)
Thus, the quantum theory of a relativistic particle constructed above
would appear to give rise to a physically unacceptable violation of
causality.

	Nevertheless, we may carry over the mathematical structure
and interpretative framework of the above theory of a relativistic
particle to our minisuperspace models for quantum gravity. When
we do so, the ``causality violation" of the theory no longer poses a
physical difficulty, since even classical trajectories in minisuperspace
do not respect the light cone structure of the DeWitt metric,
$G_{AB}$, appearing in eq.(\ref{wdw2}). Furthermore, it appears that
the other serious deficiency of the theory of the relativistic
particle -- namely, the lack of a natural choice of $\mu$ -- also can
be overcome: The metric $G_{AB}$ is invariant under translations of
the timelike coordinate $\alpha$, and the potential terms in
eq.(\ref{wdw2}) vanish asymptotically as
$\alpha \rightarrow - \infty$. Thus, it should be possible \cite{ft3}
to obtain a Hilbert space structure on the solutions to
eq.(\ref{wdw2}) in a natural way by choosing the $\mu$ associated
with this asymptotic symmetry, i.e., by choosing ${\cal H}$ to be the
subspace of solutions which asymptotically
oscillate with positive frequency \cite{ha} with
respect to $\alpha$ as $\alpha \rightarrow - \infty$.

	It should be emphasized that this asymptotic symmetry of the
Wheeler-DeWitt equation used to define ${\cal H}$ holds much more
generally: The translations in the $\alpha$-direction correspond to
scale transformations of the spatial geometry, which is a timelike
conformal isometry of all of the inverse DeWitt metrics on
full superspace \cite{ku2}. Similarly, the vanishing of the potential
as $\alpha \rightarrow - \infty$ also holds on full superspace. Thus,
our prescription for the
construction of ${\cal H}$ is not special to the particular class
of models considered here.

	The nature of the quantum theory we have just constructed
should be noted: Given any state $\Psi \in {\cal H}$, one is free to
specify any Cauchy surface, ${\cal C}$, in minisuperspace. (This
specification of ${\cal C}$ plays precisely the role of a choice of
``time" in ordinary quantum theory.) For the given state $\Psi$ at
``time" ${\cal C}$, one may then predict the probabilities for the
remaining metric variables or their conjugate momentum. In
particular, the surfaces of constant $\alpha$ are Cauchy
surfaces -- which, in fact, are naturally picked out as being
orthogonal to the Killing field used in the construction of ${\cal H}$.
If we make this choice, then we are free to specify any value of
``volume of the universe" (i.e., $\alpha$) which we wish to consider.
For any given state $\Psi \in {\cal H}$, the theory will then tell us the
probabilities for the various possible values of the conformal metric
or the trace-free extrinsic curvature at that value of $\alpha$.
Note that -- in contrast to most
other approaches to quantum cosmology -- there is a well defined
Hilbert space structure on states and well defined rules for
calculating probabilities of the above observables, but there does not
appear to be any ``preferred state" in ${\cal H}$ picked out by the
structure of the theory.

	In order to physically
interpret the theory in terms of the perceptions of
observers making measurements, it seems necessary to make
an identification of the mathematical quantity playing the role of
``time" in the theory -- namely, the choice of ${\cal C}$ -- with the
``time" as perceived by observers. This does not appear to
lead to any blatant contradictions in the context of the models
considered here, but the issue of how this identification might
generalize to the full theory (where spatial homogeneity is not
enforced) remains open (see below) and probably poses the most
significant challenge to the viability of this approach with regard to
obtaining a sensible interpretation of the full theory.

	For definiteness in our discussion, let us choose the Cauchy
surfaces of constant $\alpha$ as our specification of ``time". It should
be emphasized that within the context of this theory, it does {\it not}
make sense to ask whether the universe ``ever" achieves a given
value of $\alpha$
any more than it would make sense in ordinary quantum
mechanics to ask if a particle ever achieves a given value of time. In
effect, the identification of $\alpha$ with perceived time as proposed
above builds into the theory the expansion of the universe ``forever".
It is interesting to ask how this theory would describe Bianchi IX
cosmologies, where, in the classical theory, recollapse always occurs
\cite{lw}. (Classically, in all other Bianchi models the universe
always expands forever.) The answer is that, for any classical Bianchi
IX solution, there appears to be no difficulty in constructing a
quantum state, $\Psi$, which well approximates this classical solution
during the expanding phase in the sense that the probability
distributions for $(\beta_+, \beta_-, \phi)$ and their conjugate
momenta as functions of $\alpha$ are sharply peaked around the
values taken by the given classical solution during its expanding
phase \cite{ft4}. However, when $\alpha$ is chosen of the order of,
or larger than, the maximum value $\alpha_{max}$ achieved by the
classical solution, the behavior of the state $\Psi$ becomes
nonclassical, and, indeed, the dynamical variables
$(\beta_+, \beta_-, \phi)$ and their conjugate momenta
asymptotically become
independent of  $\alpha$ at large $\alpha$.
Further discussion of this nonclassical
behavior will be given elsewhere \cite{hw}.

	Note that the momentum, $p_{\alpha}$, conjugate to
$\alpha$ -- which measures the expansion rate of the universe in the
classical theory -- is {\it not} among the list of quantum observables
automatically defined in our theory. ($p_{\alpha}$ is, of course, the
analog of energy in the theory of the relativistic particle.) However,
${p_{\alpha}}$ can be defined by means of the Hamiltonian
constraint equation (\ref{sh}), i.e.,
\begin{equation}
{p_{\alpha}}^2 = {p_{\beta_+}}^2 +
{p_{\beta_-}}^2 + {p_{\phi}}^2 +
\exp(4\alpha) V_{\beta}(\beta_+,\beta_-) +
\exp(6\alpha) V_{\phi} (\phi)
\label{pa}
\end{equation}
since the operators appearing on the right side of this equation are
all well defined at any ``time" $\alpha$. For the Bianchi IX models,
$V_{\beta}$ can be negative, so ${p_{\alpha}}^2$ need not be
positive-definite. This corresponds to the existence of nonclassical
behavior as discussed in the preceeding paragraph. The subspace of
states associated with negative eigenvalues of ${p_{\alpha}}^2$ at
any ``time" $\alpha$ thus may be viewed as the ``nonclassical sector"
of ${\cal H}$ at that ``time". On the ``classical sector" of ${\cal H}$
(which comprises all of ${\cal H}$ except in the Bianchi IX models),
we may define $p_{\alpha}$ by taking the square root of
eq.(\ref{pa}). There exist many square roots of a positive self-adjoint
operator, but it seems most natural (and probably essential for
consistency with our interpretative remarks) to choose the positive
square root in this case. This choice implies that the universe must
be expanding whenever it can be described classically. Note that
$p_{\alpha}$ then has the interpretation of yielding the relationship
between the rate of change of
``Heraclitian time variable", $\alpha$, and that of the ``time"
registered on physical clocks.

	It should be emphasized that the above choice of positive
square root for $p_{\alpha}$ -- as well as our interpretative remarks
and our construction of ${\cal H}$ -- build ``arrows of time" into the
theory in a fundamental way. Our particular
choice of direction of these ``arrows"
was based upon the fact that the universe is observed to be expanding.
No mathematical difficulties would arise if we were to reverse the
``arrows" by identifying ``forward in perceived time" with ``decreasing
$\alpha$" and, correspondingly, were to choose the negative square
root for $p_{\alpha}$. However, it should be emphasized that,
even if one wished to do so, it would
seem difficult to restructure the theory so as to eliminate the
presence of any
``arrows". In particular, although other choices of $p_{\alpha}$ could
be made, there does not appear to be any natural choice of
``half-positive, half-negative" square root of eq.(\ref{pa}), as
presumably would be needed to obtain a ``time-symmetric" theory.

	Undoubtedly, the most crucial issue regarding all of the above
ideas and proposals is the extent to which they can be generalized to
the full superspace case. In order to do this, the following obstacles
must be overcome: (1) An analog of a Cauchy surface for the
Wheeler-DeWitt equations on superspace must be found. (2) A
symplectic product on the solutions to the Wheeler-DeWitt equations
(whose value is independent of choice of Cauchy surface) must be
identified. (3) A suitable subspace of solutions must be chosen to
serve as the Hilbert space of states, ${\cal H}$. (4) Our construction of
position and momentum operators must be generalized. (5) Finally,
in order to obtain an interpretation of the theory in terms of local
measurements made by observers, it would appear necessary to
suitably identify the variables defining Cauchy surfaces in
superspace with the ``perceived time" of local observers.

	It will, of course, be necessary to confront the
difficult issues of regularization and renormalization of the theory
in order to analyze the above issues in a mathematically
satisfactory way.
However, it would appear that issues (1)-(3) and perhaps (5) can be
at least partially investigated \cite{b} without getting deeply involved in
regularization issues.
The Wheeler-DeWitt equations comprise an infinite
family (one for each choice of lapse function, $N$), and one would
expect a ``Cauchy surface" in superspace to have co-dimension equal
to the number of Wheeler DeWitt equations. Thus, something of the
nature of a cross-section of the conformal geometries on superspace
would appear to be a good candidate for a Cauchy surface (with the
conformal factor thus playing the role of ``time", i.e., labeling the
Cauchy surfaces). However, it is not at all clear whether this
suggestions works in detail and what, if any,
additional conditions might
need to be imposed upon the cross-section. In this regard, it should
be noted that the only obvious metrical structure present on
superspace is the infinite family of
DeWitt inverse supermetrics (which, in general at least, are
degenerate \cite{fh}), but it is not immediately
clear whether or how this structure might be used to
obtain an analog of ``Cauchy surfaces".
The symplectic product should have the basic structure of
the DeWitt product (see eq.(5.19) of \cite{dw}),
but it is far from clear
exactly what form it would take when expressed as an integral
over a Cauchy surface in superspace. (Unfortunately, it also would appear
that regularization issues will play a prominent role in defining
the symplectic structure.) If the
solutions to the Wheeler-DeWitt equation behave in an ``ultralocal"
manner \cite{pi} as the metric is scaled to zero (corresponding to the limit
$\alpha \rightarrow - \infty$ in our minsuperspace models) then it
should be possible to define ${\cal H}$
as the subspace of solutions which
asympotically oscillate with positive frequency with respect to
independent conformal scalings at each point of space.

	The above issues are presently under investigation \cite{hw}.

	I wish to thank A. Higuchi for reading this manuscript
and for many useful discussions. I particularly wish to thank
Karel Kuchar for several extremely
helpful discussions.
This research was supported in part by
National Science Foundation
grant PHY-9220644 to the University of Chicago.


\begin{thebibliography}{99}
\bibitem{ft1} Here we employ the ``traditional" ADM canonical
variables. It is essential for our approach that the canonical variables
be chosen so that the constraints be at most quadratic in the
momenta, but otherwise one should be free to consider other choices,
such as the ``Ashtekar variables".
\bibitem{ft2} This difficulty in the theory of a relativistic particle
corresponds to the difficulty in defining a meaningful notion of
``particles" in quantum field theory. However, this does not pose any
problem for the formulation of quantum field theory, which does not
require that any notion of ``particles' be defined; see \cite{wa1} for
further discussion.
\bibitem{ft3} Since the potential terms appearing in eq.(\ref{wdw2})
typically will be unbounded at large ``spatial coordinates", the
manner in which they vanish asymptotically as
$\alpha \rightarrow - \infty$ is non-uniform, so it is not obvious
that the prescription for defining $\mu$ always can be applied.
Indeed, for vacuum Bianchi VIII and IX models, the behavior of the
$V_{\beta}$ as $\alpha \rightarrow - \infty$ is well approximated
by ``confining walls" which move at velocity less than light, and
hence $V_{\beta}$ cannot be neglected in either the classical or
quantum dynamics. However, even in these models, when the
$\phi$ (or some other fundamental matter field)
degree of freedom is added,
the ``walls" are no longer confining (provided that $V_{\phi}$ does
not grow exponentially rapidly with $\phi$), and the asymptotic
symmetry needed to define $\mu$ should be present (see \cite{hw}
for further discussion). Thus, the viewpoint adopted here is that the
asymptotic dynamics of the vacuum Bianchi VIII and IX models is of
a very special character, and that the generic asymptotic dynamics is
``velocity dominated" as $\alpha \rightarrow - \infty$.
\bibitem{ft4} Similarly, one also could construct a state $\Psi '$
which approximates the the classical solution during its contracting
phase, and one also could take superpositions of these states.
\end{thebibliography}
\end{document}